\documentclass[aip,rsi,reprint]{revtex4-1}
\usepackage{textcomp}
\usepackage{graphicx}
\usepackage{epstopdf}
\usepackage{caption}
\usepackage{subcaption}

\begin{document}

\title{Micro-Faraday cup matrix detector for ion beam measurements in fusion plasmas}

\author{D. I. R\'efy}\email{refy.daniel@wigner.mta.hu}
\affiliation{Wigner Research Centre for Physics, Plasma Physics Department, 1121 Budapest, XII. Konkoly Thege Mikl\'os \'ut 29-33., Hungary}

\author{S. Zoletnik}
\affiliation{Wigner Research Centre for Physics, Plasma Physics Department, 1121 Budapest, XII. Konkoly Thege Mikl\'os \'ut 29-33., Hungary}

\author{D. Dunai}
\affiliation{Wigner Research Centre for Physics, Plasma Physics Department, 1121 Budapest, XII. Konkoly Thege Mikl\'os \'ut 29-33., Hungary}

\author{G. Anda}
\affiliation{Wigner Research Centre for Physics, Plasma Physics Department, 1121 Budapest, XII. Konkoly Thege Mikl\'os \'ut 29-33., Hungary}

\author{M. Lampert}
\affiliation{Wigner Research Centre for Physics, Plasma Physics Department, 1121 Budapest, XII. Konkoly Thege Mikl\'os \'ut 29-33., Hungary}

\author{S. Heged\H{u}s}
\affiliation{Wigner Research Centre for Physics, Plasma Physics Department, 1121 Budapest, XII. Konkoly Thege Mikl\'os \'ut 29-33., Hungary}

\author{D. Nagy}
\affiliation{Wigner Research Centre for Physics, Plasma Physics Department, 1121 Budapest, XII. Konkoly Thege Mikl\'os \'ut 29-33., Hungary}

\author{M. Pal\'ankai}
\affiliation{Wigner Research Centre for Physics, Plasma Physics Department, 1121 Budapest, XII. Konkoly Thege Mikl\'os \'ut 29-33., Hungary}

\author{J. K\'adi}
\affiliation{Wigner Research Centre for Physics, Plasma Physics Department, 1121 Budapest, XII. Konkoly Thege Mikl\'os \'ut 29-33., Hungary}

\author{B. Lesk\'o}
\affiliation{Wigner Research Centre for Physics, Plasma Physics Department, 1121 Budapest, XII. Konkoly Thege Mikl\'os \'ut 29-33., Hungary}

\author{M. Aradi}
\affiliation{Fusion@ÖAW, Institute of Theoretical and Computational Physics, Graz University of Technology, Petersgasse 16, A–8010 Graz, Austria}

\author{P. Hacek}
\affiliation{Institute of Plasma Physics of the CAS, Tokamak Department, 182 00, Prague, Czech Republic}

\author{V. Weinzettl}
\affiliation{Institute of Plasma Physics of the CAS, Tokamak Department, 182 00, Prague, Czech Republic}

\begin{abstract}

Atomic Beam Probe (ABP) is an extension of the routinely used Beam Emission Spectroscopy (BES) diagnostic for plasma edge current fluctuation measurement at magnetically confined plasmas. Beam atoms ionized by the plasma are directed to a curved trajectory by the magnetic field and may be detected close to the wall of the device. The arrival location and current distribution of the ions carry information about the plasma current distribution, the density profile and the electric potential in the plasma edge. This paper describes a micro-Faraday cup matrix detector for the measurement of the few microampere ion current distribution close to the plasma edge. The device implements a shallow Faraday cup matrix, produced by printed-circuit board technology. Secondary electrons induced by the plasma radiation and the ion bombardment are basically confined into the cups by the tokamak magnetic field. Additionally, a double mask is installed in the  front face to limit ion influx into the cups and supplement secondary electron suppression. The setup was tested in detail using a Lithium ion beam in the laboratory. Switching time, cross talk and fluctuation sensitivity test results in the lab setup are presented, along with the detector setup to be installed at the COMPASS tokamak.
\end{abstract}

\maketitle

\section{Introduction\label{Sect.introduction}}
The ELMy H-mode operation regime of magnetically confined fusion plasmas is being explored for several decades\cite{Wagner2007} and it is still in the focus of fusion research as it considered to be the main operating regime of a future fusion reactor. Alongside the benefits of H-mode in terms of improved energy  confinement compared to L-mode operation, the so called Edge Localized Modes (ELMs) are of severe concern for a future reactor.  They are an important element to extract impurities from the plasma but at the same time they periodically expel up to $20\%$ of the total plasma energy within a millisecond time scale\cite{Doyle2007}. Such a high power load can damage the plasma facing components of the machine. 

The understanding of the physics background of the ELM triggering mechanism is indispensable in order to control the ELMs and to mitigate their effect. The key physics quantities for the peeling-ballooning instability which can describe the type-I (large) ELMs are the plasma edge pressure gradient and current density\cite{Snyder2002}. The pressure gradient can be measured by several techniques which are available on numerous machines e.g. Alkali-BES\cite{Refy2018} and reflectometer\cite{Wang2004}$^{,}$\cite{Meneses2006} for electron density profile measurement, electron cyclotron emission\cite{Classen2010}$^{,}$\cite{Luna2004} for electron temperature, charge exchange recombination spectrosopy\cite{Isler1994} for both the ion temperature and ion density, Thomson scattering\cite{Pasqualotto2004} for both the electron temperature and the electron density measurement. On the other hand there are only limited possibilities for the edge current density measurement, therefore it is not measured routinely.

A trajectory of a charged-particle beam passing through a magnetically confined plasma is determined by the energy, charge state and mass of the given particle, the magnetic field structure and the electric potential distribution. The measurement of the trajectory of a monoenergetic ion beam thus enables  characterization of the magnetic field and potential. Several techniques have been proposed both for the diagnostic beam production and the ion beam trajectory detection. 

The Heavy Ion Beam Probe (HIBP) technique \cite{Jobes1970} utilizes a primary beam of singly charged ions of a large mass species ($Cs^{+}$ typically), which may undergo ionization after injection into the plasma due to collisions with the plasma particles. The doubly charged ions are separated from the primary beam at the point of second ionization, follow a path defined by the magnetic field, and exit the confined region at one point, reaching the wall of the machine. The standard HIBP technique is based on the detection of the spatial \cite{Beckstead1997} and energy distribution of the leaving ions which reflects the variation of the plasma current, electron density and plasma potential mostly at the second ionization location. Additionial  measurements have also been proposed for secondary ion beam emission imaging \cite{Demers2003} and beam velocity direction measurement\cite{Fimognari2016}. Variations of the HIPB technique indeed allow excellent measurements but serious limitations arise from the necessary high (up to MeV) beam energy and limited access to the plasma. 

A proposed version of HIBP is called the Laser-accelerated Ion-beam Trace Probe (LITP) \cite{Yang2014}. It intends to replace the electrostatic particle acceleration method by laser ablation. In this case a pulsed beam of MeV ions with high energy spread and multiple charge states is obtained. The concept intends to measure the spatial distribution of ions at the wall of the fusion machine which enables the reconstruction of the radial electric field, electron density\cite{Chen2014} and the poloidal magnetic field \cite{Yang2016}.

The Atomic Beam Probe\cite{Berta2013} (ABP) technique to be discussed in this paper is an extension of the routinely used Alkali (typically Lithium or Sodium) Beam Emission Spectroscopy diagnostic\cite{Zoletnik2018}. The ion source of the system is replaceable on a day timescale, thus the ion species can be varied flexibly. Lithium and Sodium is used routinely and other heavier species (Rb, Cs) are also possible. ABP intends to measure the spatial distribution of an ion beam originating from the atomic beam after ionization in the plasma. The ion beam path is affected by the magnetic field and the electric potential thus the ion beam shape enables measurement of these quantities. The beam current depends on the plasma density, therefore information can be obtained on this quantity as well. Unlike in the case of HIBP, the ion energy cannot be measured with the required precision inside the fusion device, therefore the diagnostic relies solely on the ion beam current distribution measurement. On the other hand, the technique is a simple extension of a standard diagnostic and potentially offers microsecond-scale temporal resolution. 

For detection of the ion beam distribution a collection plate matrix was proposed and tested in the COMPASS tokamak\cite{Hacek2018}. Results showed that several improvements are necessary to be able to extend the operation space towards standard H-mode scenarios. The detector has to be first tested in laboratory environment to quantify the performance and to be able to interpret the measurements in a tokamak environment. This paper presents the design and laboratory testing of a new ABP detector. It has to be noted that an alternative scintillator detector concept has also been proposed\cite{Sharma2017} which offers better spatial resolution but more limitation on geometry.

This paper is organized as follows: the working principle of the ABP is described briefly in section \ref{Sect.principle}. The detector head for the ion current measurement is presented in section \ref{Sect.detector}. The setup for the laboratory experiment is detailed in section \ref{Sect.labexp}  followed by the measurement results in section \ref{Sect.labexpresults}. The ABP setup for the tokamak environment is described in section \ref{Sect.tokexp}, and the paper is summarized in section \ref{sect:summary}.

%

\section{ABP working principle\label{Sect.principle}}

The ABP is an extension of the alkali BES diagnostic described in detail in ref\cite{Zoletnik2018}.  The diagnostic is routinely used for electron density profile measurement at several plasma experiments, and works as follows. An accelerated atomic beam is injected into the plasma, where the beam atoms are excited and ionized by plasma particles. The ionization process results in a gradual loss of the atoms in the beam. The ionization rate is such that the beam can penetrate only the edge of the plasma, thus Li-BES systems are used for electron density profile and fluctuation measurement of the outer plasma regions only, namely the plasma edge and scrape off layer (SOL). Spontaneous de-excitation of the beam atoms results in a characteristic photon emission which can be detected through an optical system. The distribution of the light emission along the beam (light profile) can be measured by a detector system, from which the electron density distribution (density profile) can be calculated \cite{Schweinzer1992}$^,$\cite{Fischer2008}.

\begin{figure}
\begin{centering}
\includegraphics[width=\linewidth]{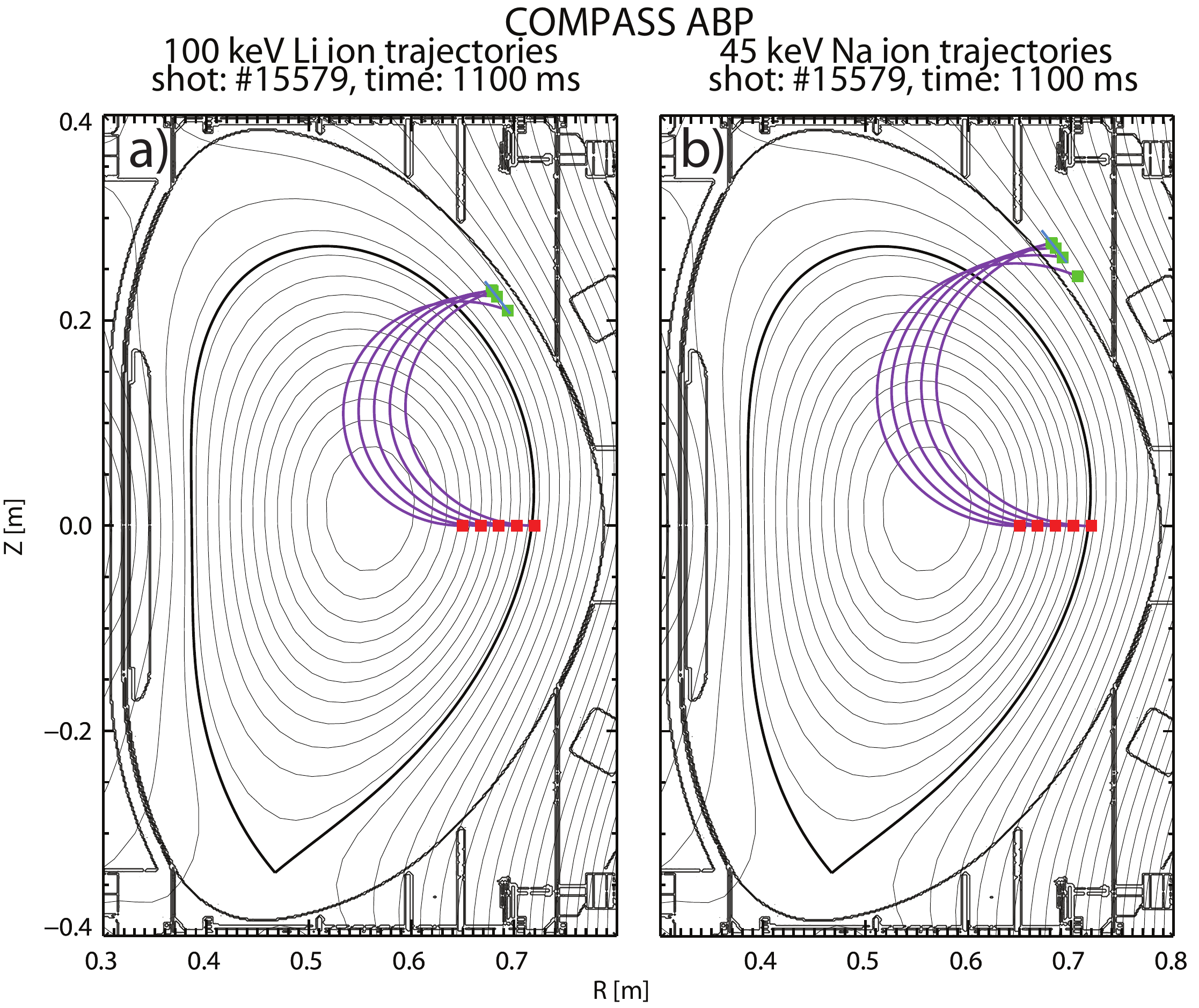} 
\par\end{centering}
\caption{100 keV Lithium (a) and 45 keV Sodium (b) ion beam trajectories (solid purple lines) along with the ionization location (red squares), the detector location (solid blue line) and the detector plane impact location (green squares) are shown. The poloidal cross section of the COMPASS vacuum vessel and the magnetic surfaces with the highlighted last closed flux surface are also indicated. (shot: \#15579, time: 1100 ms) \label{fig:polcut}}
\end{figure}

The ions are deflected from the beam by the magnetic field, and may reach the wall of the machine. Figure \ref{fig:polcut} shows the modelled path of 100 keV Lithium (a) and 45 keV Sodium (b) ions in the COMPASS alkali beam diagnostic\cite{Anda2016}. Ions from the beam injected from the low field side (LFS) midplane reach the wall of the machine approximately 22-27 cm above the midplane on the LFS. Here the ion detector can be placed into a port. The red squares indicate the location of the ionization, while the green squares show the intersection of the ion trajectories and the detector surface plane. The ion trajectories are deflected in the toroidal direction (out of the plane of the drawing) by the poloidal magnetic field resulting from the plasma current. This toroidal displacement needs to be measured by the detector. Ions starting at different locations in the plasma hit the detector at different elevation and different toroidal displacement, therefore a two-dimensional measurement is desirable. Besides the toroidal deflection, a change in the vertical detection position is caused by the plasma potential change or toroidal magnetic field change. Intensity modulations are caused by plasma density fluctuations at the ionization point and outside. Disentangling these effects is not straightforward and will be the subject of other publications.

\section{Detector setup\label{Sect.detector}}

\subsection{Detector plate\label{subs.head}}

The detector is considered as a two-dimensional matrix of ion collector metal plates. The simulation of the ion trajectories shows that the detector has to be placed inside the tokamak vessel, typically in a few centimeter distance from the last closed flux surface (LCFS), facing the plasma. This predetermines high - mostly ultra violet and X-ray - radiation level at the detector which causes secondary electron production on the detector surface. The secondary electron current can be significantly higher than the ion current, therefore the detector design should minimize secondary electron effects.
Secondary electrons leave the metal surface with few 100 eV energy. In the few Tesla magnetic field of the tokamak the Larmor radius of these electrons is below 100 micron. If the detector plate is parallel to the tokamak magnetic field a few hundred micron deep cup can prevent them from leaving.  Due to variable magnetic configuration a fully parallel magnetic field cannot be ensured on the detector but the deviation can be kept below 20 degrees. In this case a factor of 3 toroidal width/depth ratio of the cup is sufficient to confine electrons in any case. 

\begin{figure}
\begin{centering}
\includegraphics[width=\linewidth]{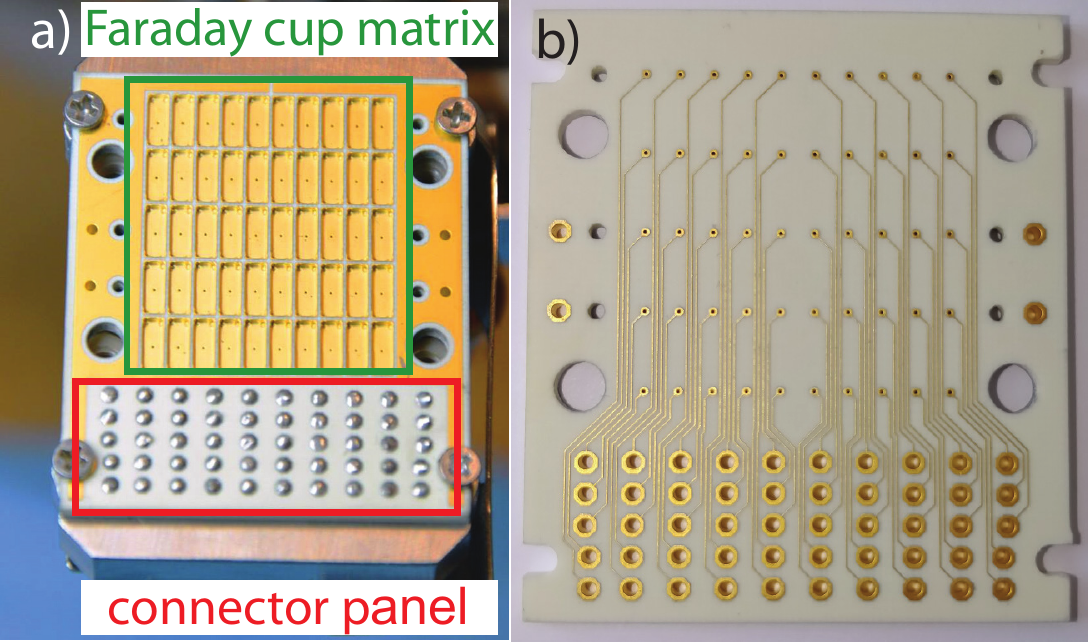} 
\par\end{centering}
\caption{Front view (a) and back view (b) of the detector panel. The $5 \times 10$ faraday cup matrix and the connector panel below can be seen in the front view, while the wiring between the connectors and the plates on the back view. \label{fig:detectorhead}}
\end{figure}

The cup size is determined the following way. According to computer simulation of the COMPASS measurement setup, the ion beam toroidal shift on the detector in response to an edge current perturbation is 0.1-1 mm/kA. In earlier experiments a Lithium beam reduced to 5 mm still gave good signal\cite{Hacek2018}, therefore we considered this beam diameter. We intend to detect at least 1 mm beam movement, thus the 5 mm beam needs to be resolved to a few measurement channels. As a consequence, the Faraday cup toroidal width was set to 2 mm. In the vertical direction much less resolution is needed, therefore the height was set to 5 mm. The total detector size is limited by the COMPASS port width, therefore a 5x10 (row$\times$column) matrix was possible.  The final dimensions of each Faraday cup are $1.8\times 4.8 \times 0.8$ mm (toroidal width, height, depth), the distance between the cups is 0.4 mm in both directions. Figure \ref{fig:detectorhead} shows the detector board from the front (a) and the back (b). Each cup is connected to a wire connection point at the lower edge of the detector panel which is shielded from the plasma.

\begin{figure}
\begin{centering}
\includegraphics[width=\linewidth]{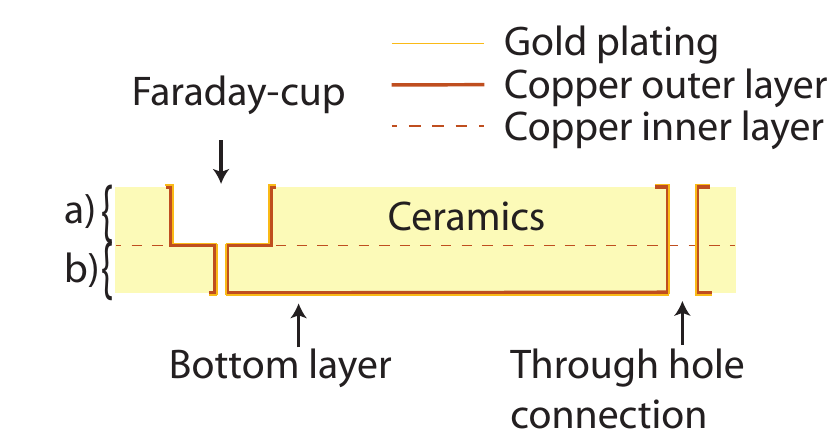} 
\par\end{centering}
\caption{Schematic drawing of the detector layer setup. a) Upper layer with one copper-plated side, b) Bottom layer with double sided copper-plating. The two layers are laminated together and then electroplated to form the detector setup.
 \label{fig:circuit}}
\end{figure}

The detector was manufactured using standard PCB (Printed Circuit Board) technology, however, PCB production is not utilized to produce three-dimensional structures, which was a necessity for the Faraday-cups in the detector. Another difficulty had to be overcome due to the possible high heat loads on the detector board, which excluded the use of standard FR4 fiberglass material. A possible candidate for the PCB was found in the high frequency radio industry, which utilize ceramics as the base material of the circuit board (RO4350B)\cite{RO4000}. Ceramics can withstand the high heat loads during Atomic Beam Probe measurements since the material meets the flame-retardant standards of Restriction of Hazardous Substances (RoHS), and the PCB can be manufactured using standard FR4 production line. The other difficulty of producing a three-dimensional structure into the PCB was overcome by an advanced layer setup. Three layers were prepared for the setup, as can be seen in Figure \ref{fig:circuit}. The first layer is used for the routing of the cables from the Faraday cups to the electrical connection. The second layer provides the bottom part of the Faraday-cups. The third layer is first milled for the slits of the Faraday-cups. In the next step, the first two and the third layer are laminated together, which is followed by copper metallization, which connects the bottom of the Faraday-cups to the top layer and forms the side of the cups, as well. As a last step, the wires are gold plated finalizing the detector board. The final detector was manufactured by Hungarian team of MMAB Group \cite{MMAB}.

\subsection{Detector mask}\label{subs.mask}

\begin{figure}
\begin{centering}
\includegraphics[width=\linewidth]{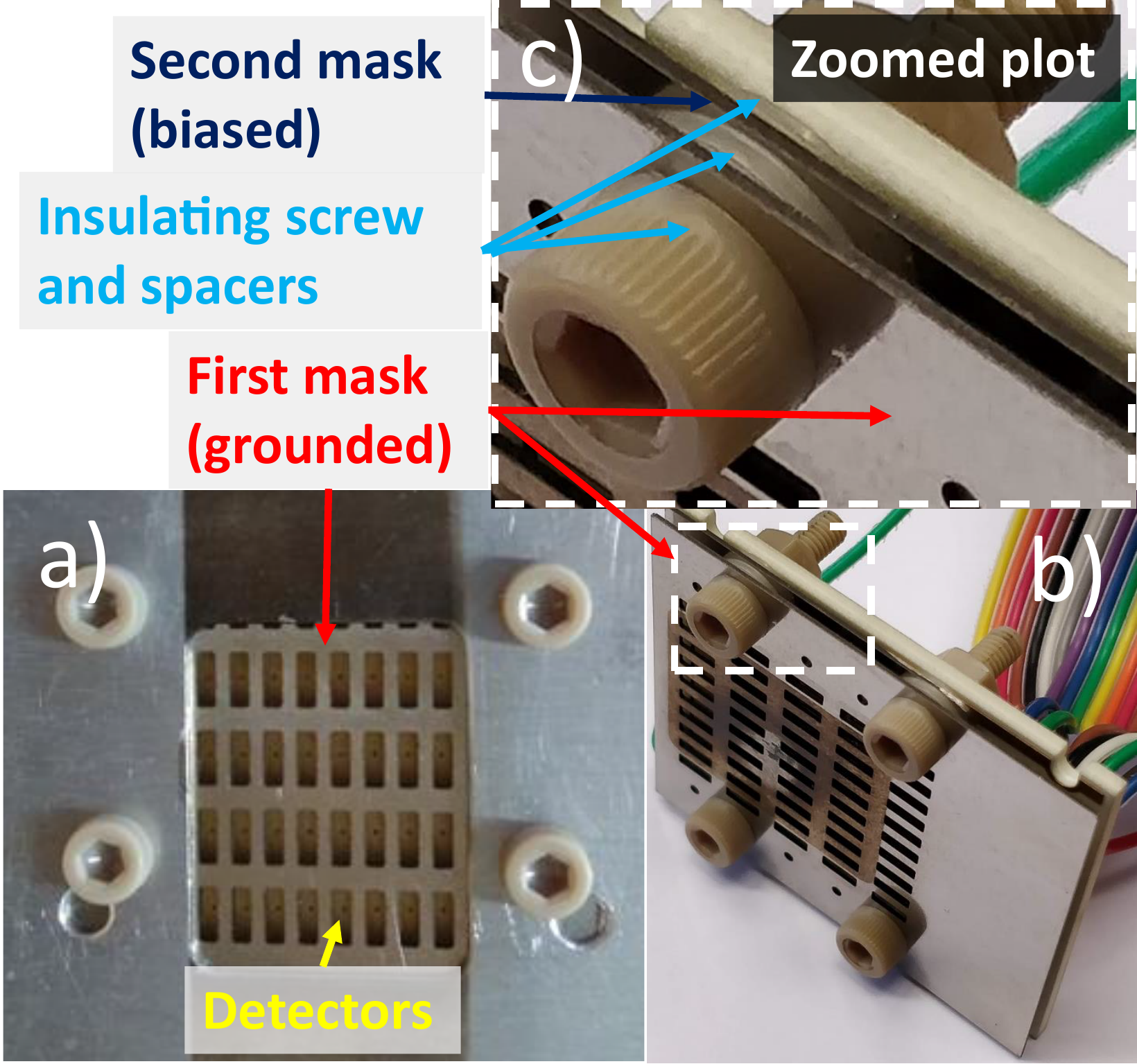} 
\par\end{centering}
\caption{Front view (a) of the detector masks installed in the laboratory setup, the detectors are visible behind, side-front view (b) of the detector head with the double mask and the cables behind, and a zoomed plot (c) showing the masks, the insulating spacers and screws. \label{fig:detectormask}}
\end{figure}

Previous measurements\cite{Hacek2018} showed that electrons and ions as well as UV radiation reaching any metallic surface induce secondary electrons which make the measurement hard to interpret. The detector plate has to be masked so that the ions can only reach the detector at the Faraday cups. On the other hand any secondary electrons generated at the Faraday cup edges e.g. by stray UV radiation should be suppressed by an external electric field while the detector plate and its front face have to be on the ground potential.  
To fulfill these requirements, a double mask is placed in front of the detector, as indicated in front and front-side and zoomed views in Figure \ref{fig:detectormask} (a), (b) and (c). The openings in the mask are 1.2 mm~$\times$~3.6 mm, that is somewhat smaller then the Faraday cup size. The masks are separated electrically from each other and from the detector as well by insulating washers. The spatial separation is 1 mm between each element. The mask closer to the Faraday cup can be either biased or grounded, while the one further facing the plasma is grounded to prevent collection charges from the plasma.

\begin{figure}
\begin{centering}
\includegraphics[width=\linewidth]{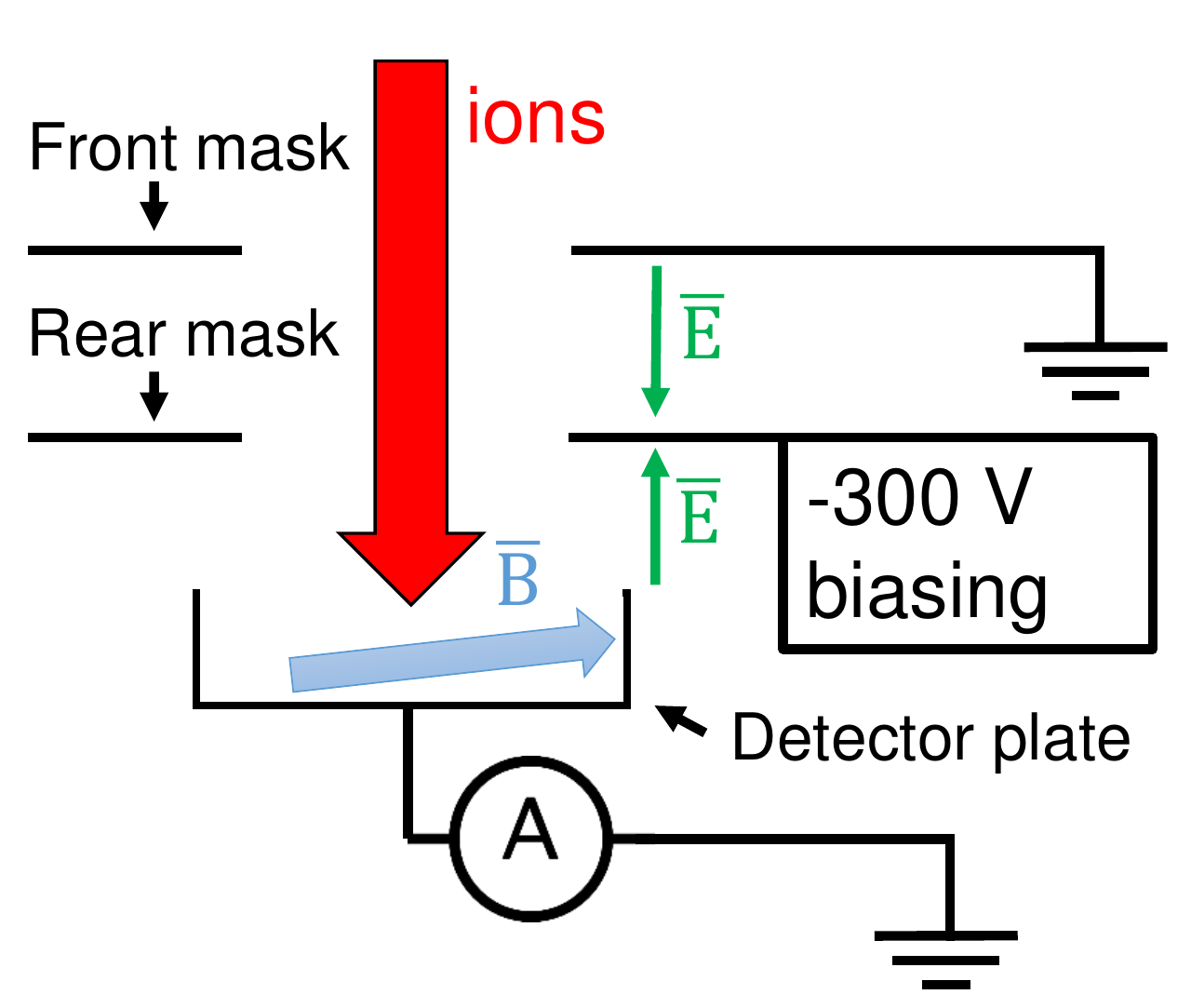} 
\par\end{centering}
\caption{Double mask concept: Ions are following straight trajectory on the mm scale, and are indicated by the red arrow. The front grounded mask prevents ion impact on surfaces between the detectors. Direct ion impact on the detector surface induces secondary electrons which are prevented from leaving the cup either due to their small Larmor radius and by hitting the wall of the cup after following the magnetic field lines (indicated by the blue arrow). Electrons generated on the cups edges are pushed back by the electric field produced by the biased rear mask (indicated by the green arrows). \label{fig:detectordraft}}
\end{figure}

Figure \ref{fig:detectordraft} shows a sketch of the typical ion impact scenario, looking from the top at the detectors, quasi perpendicular with the local magnetic field. The Larmor radius of a 100 keV Lithium ion is in the order of 10 cm in 1 T magnetic field, and its trajectory is considered to be straight on the mm length scale as indicated with the red arrow. The Larmor radius of a few 100 eV electron in 1 T magnetic field is in the order of 10 $\mu$m. As the magnetic field is close to parallel to the cup surface, the electrons are primarily prevented from leaving the cup by their small Larmor radius. Electrons which might leave the surface will travel along the magnetic field line, indicated with the blue arrow, and will hit the side of the cup. Should any electron be generated on edge of the cups they are pushed back by the electric field produced by the biased rear mask indicated by the green arrows.

\section{Laboratory experiment setup\label{Sect.labexp}}

\begin{figure*}
\begin{centering}
\includegraphics[width=\linewidth]{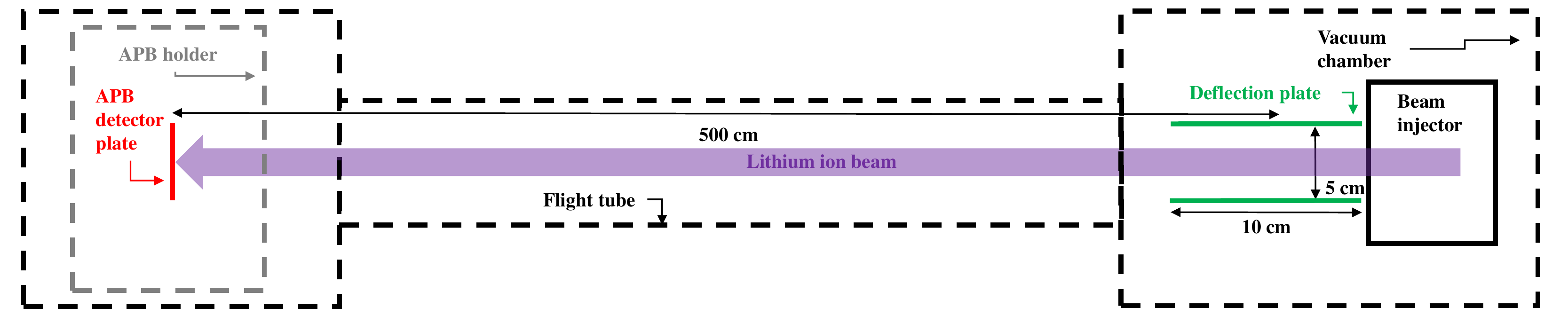} 
\par\end{centering}
\caption{Laboratory diagnostic setup with ABP placed far from the ion source \label{fig:diagsetupblock3}}
\end{figure*}

A laboratory experiment was conducted to verify secondary electron suppression, temporal resolution and crosstalk of the detector. A 30 kV Li-beam injector was used as the ion source in a setup shown in Figure \ref{fig:diagsetupblock3}.

\subsection{Lithium beam injector\label{Sect.injector}}

The ion beam required for the measurement is produced by a simplified version of the Li-beam injector described in Ref.~\cite{Anda2016}. The  approximately 1 mA ion beam is extracted with 3 kV and accelerated with 27 kV from a thermionic emitter with a 2 stage ion optic. The beam is injected through a flight tube, and chopped (deflected so that the beam cannot reach the detector) with a pair of deflection plates.  One is grounded, while the other is connected to a fast switch which can change between the potential set on a power supply up to 1 kV and the ground with up to 500 kHz switching frequency. The distance between the detector and the chopper is about 5 m. The beam has about 2 cm FWHM at the detector, thus covering all channels when the beam is not modulated. When the beam modulation is done with 900 V it results in 15 cm deflection at the detector plate, that is the beam is completely off the detector. Note that the beam conditions, such as emitter temperature, emitter depletion, extraction voltage, and beam current, and beam focusing accordingly can be significantly different between the measurement series to be presented in this paper; thus, the net measured current per detector is not necessarily comparable between the experiments.

\subsection{Detector holder}

\begin{figure}
\begin{centering}
\includegraphics[width=\linewidth]{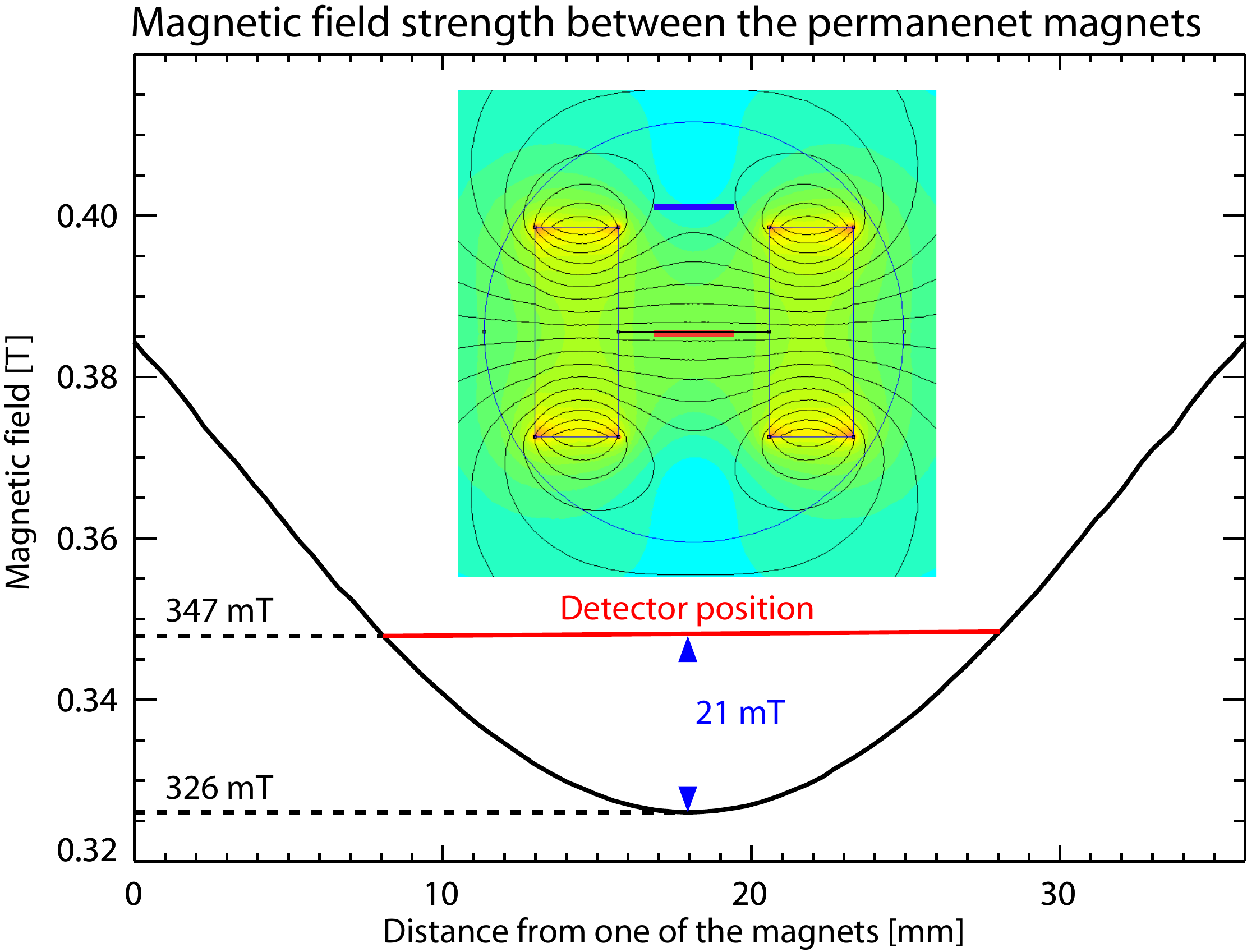} 
\par\end{centering}
\caption{Strength of the magnetic field in between the Neodymium magnets as a function of the distance from one magnet. The setup viewed from the top is shown in the contour plot, the graph shows the horizontal cut of the 2D simulation results at the middle of the setup indicated with a black line. The detector position is indicated with a red line. The detector position for the Ref.~\cite{Hacek2018} setup with less homogenous field is also indicated with a blue line \label{fig:comsol}}
\end{figure}

A detector holder was designed  to mimic the tokamak magnetic field. Two Neodymium magnets with about 0.5 T surface induction were placed on 2 sides of a holder which is hung from the vacuum flange. The detector is placed in the middle of the magnets where the field is the strongest and the most homogeneous. The setup was modelled with Finite Element Method Magnetics\cite{Meeker2018} as indicated in Figure \ref{fig:comsol}, showing a straight induction vector at the detector position (indicated with a red line), and 7$\%$ field strength variation (326 - 347 mT) between the edge and the center of the 20 mm wide detector. The detector is perpendicular to the geometrical beam line axis, however the beam is slightly deflected downwards due to the magnetic field. (30 keV, Li, 0.4 T homogenous field, 2 cm path in the field, $r_{L}$=16 cm, deflection: 2.5 mm, $7^\circ$) The major difference of the setup described in Ref.~\cite{Hacek2018} is that the magnetic field was less homogenous at the detector position which is indicated with a blue line in Figure \ref{fig:comsol}.

\begin{figure}
\begin{centering}
\includegraphics[width=\linewidth]{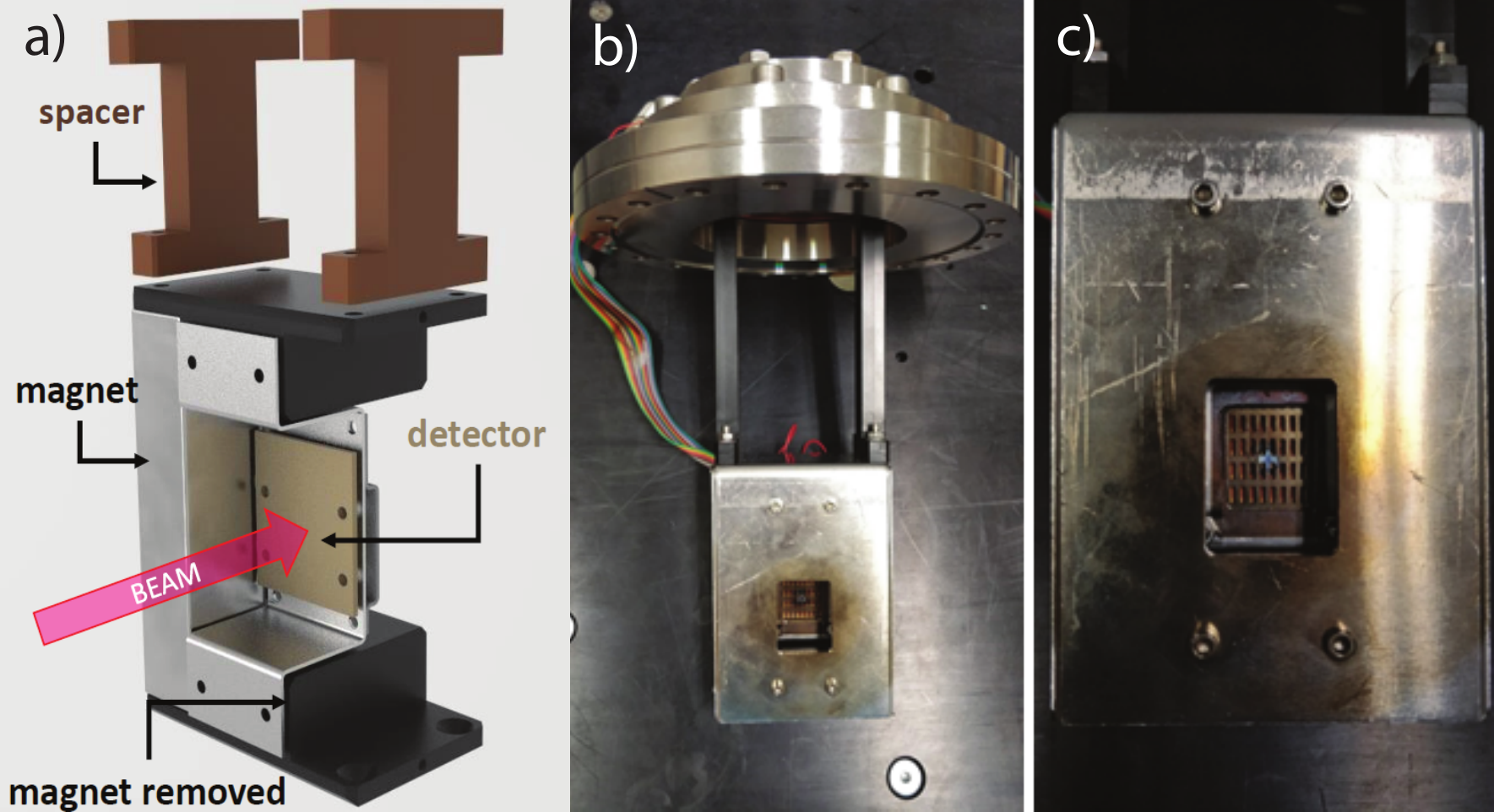} 
\par\end{centering}
\caption{CAD model with one magnet and the covering mask removed (a), full setup with the vacuum flange (b) and the detector head with the magnet holder zoomed (c). \label{fig:detectorholder}}
\end{figure}

Figure \ref{fig:detectorholder} shows the Computer-Aided Design (CAD) model with one magnet and the covering mask removed (a), the full setup with the vacuum flange (b) and the detector head with the magnet holder zoomed (c). 
The Faraday cup signals are led to the vacuum feedthrough by a ribbon cable, and the external voltage and grounding are also applied there.

\subsection{Data acquisition and control}\label{Sect.labcontrol}

The signals from 20 Faraday cups were connected through a 2 m bundle of coaxial cables to amplifiers consisting of a current sensing stage with 2 kOhm resistor and  a second stage amplifier with gain of 100. The analogue bandwidth was 1 MHz and the differential signals were digitized with 2 MHz and 14 bits resolution.  The sampling of the digitizer was synchronized to the beam modulation at the deflection plate pair.  A 1 kOhm series resistor and voltage limiting diodes were installed at the amplifier input to protect them from overcurrent from the plasma or beam. The time constant to the resistor and the cable capacitance contributes to to final analogue bandwidth of the setup. This setup is identical to the one used on the COMPASS tokamak.

Unused cups were connected to the ground. To prevent noise pickup, proper grounding turned out to be essential. Therefore, the detector setup was isolated from the beam injector and the beam flight tube and grounded from the amplifiers and digitizers.

\section{Laboratory experiment results} \label{Sect.labexpresults}

\subsection{Mask biasing test \label{subs:biasing}}

The aim of this test was to check the amount of spurious signal in the Faraday cups from secondary electrons generated by ion impact.  If the beam is injected fully perpendicularly to the detector surface all ions hit the Faraday cup bottom and secondary electrons are confined in the cup by the magnetic field. However, in the tokamak experiment the beam is not always perpendicular and also in the laboratory experiment it is deflected about 7 degrees by the magnet. This way some electrons are generated at the side of the cups close to the top and may escape. The secondary electrons are suppressed when negative voltage is applied, while extracted from the plate when the mask is positively biased. 

\begin{figure}
\begin{centering}
\includegraphics[width=\linewidth]{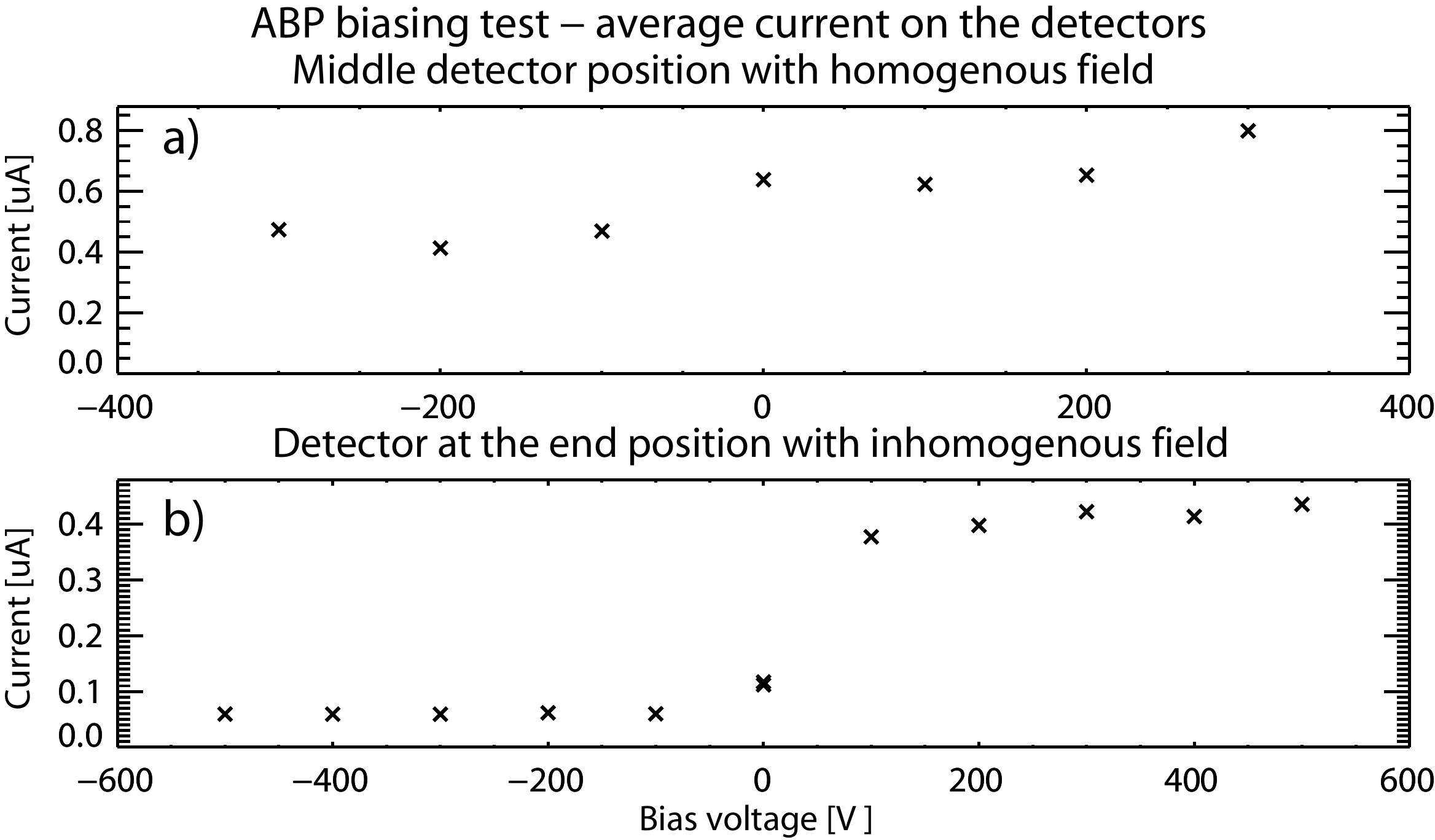} 
\par\end{centering}
\caption{Average current on the detectors as a function of biasing voltage. Detector was placed in homogenous field (a) and in inhomogeneous field (b) as indicated with the red and the blue lines respectively in Figure \ref{fig:comsol}. There is a clear decrease when negative voltage is applied.\label{fig:biastest2}}
\end{figure}

The signal level was measured on the Faraday cup plates while the biasing voltage of the mask was scanned. Figure \ref{fig:biastest2} shows the average current on the plates as a function of the biasing voltage. Figure \ref{fig:biastest2} (a) corresponds to the homogenous field setup while Figure \ref{fig:biastest2} (b) to the inhomogeneous field setup indicated with red and blue respectively in Figure \ref{fig:comsol}. The systematic effect of the biasing can be seen when negative voltage is applied. Even at -100 V the current drops from 0.6 $\mu$A to 0.45 $\mu$A for the homogenous case. The effect is more emphasized when the detector was in inhomogeneous field, since the current changes by factor 8 between the negatively and the positively biased mask cases (0.4 $\mu$A to 0.05 $\mu$A). This result confirms that the geometrical electron suppression described in section \ref{subs.mask} works reliably for small pitch angle between the detector surface and the local magnetic field, and suggests small (few hundred volts) negative biasing for the measurement. The beam current was relatively stable during each measurement series, but small deviations could occur which can explain the small deviations from the trend. The beam current changed considerably between the homogenous (0.45 $\mu$A ion current) and the inhomogeneous case (0.05 $\mu$A ion current) due to the different emitter conditions and beam focusing.

\subsection{Cross talk test \label{subs:crosstalk}}
The aim of this measurement was to characterize the cross talk between channels due to escaping secondary electrons. A special mask was installed with only one 1.2 mm $\times$ 3.6 mm window, allowing ions to reach directly only one Faraday cup.

\begin{figure}
\begin{centering}
\includegraphics[width=\linewidth]{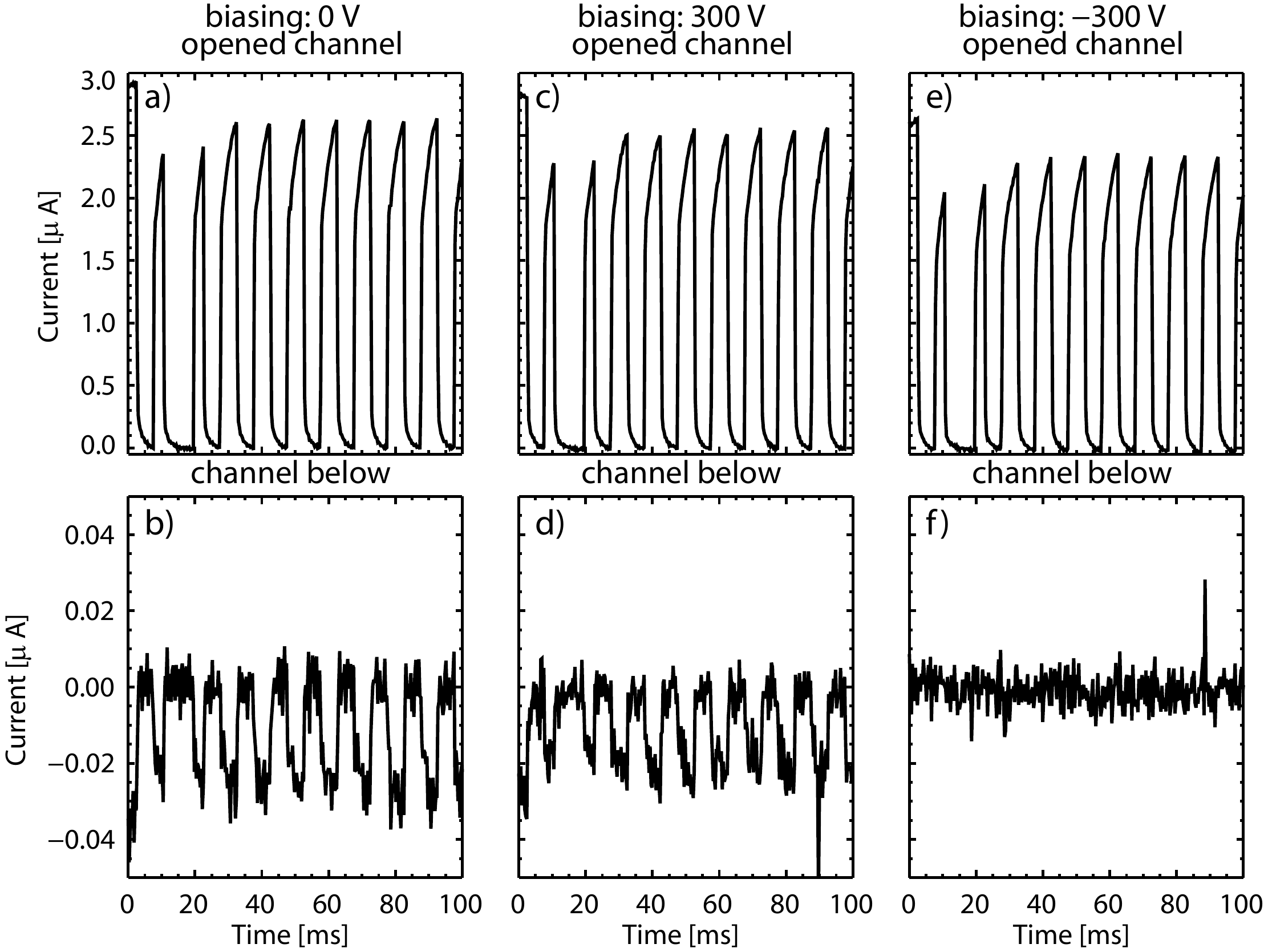} 
\par\end{centering}
\caption{Cross talk test, signals above show the signal level on the channel behind the opening on the mask while signals below show the signal level on the channel below the opening for various biasing: mask grounded (a,b), mask biased with +300 V (c,d), mask biased with -300V (e,f) \label{fig:crosstalkall}}
\end{figure}

The 30 keV, 1 mA ion beam was injected and chopped with 100 Hz.
A measurement series was done with different mask biasing: grounded, +300 V, -300 V. The raw signals with 10 $\mu$s integration time are plotted for all channels in Figure \ref{fig:crosstalkall} for the different biasing cases. The beam current changes on the 10 ms time scale since the ion beam focusing time can be several tens of milliseconds due to space charge being compensated by back flowing electrons in the beam. 
There is about 1.5\% crosstalk with the bottom channel relative to the opened one with the unbiased and the positively biased cases (consider different vertical axis scale), while no cross talk for the negatively biased case. It is also visible that the signal on the bottom channel is in counter phase with the reference channel indicating the effect of the secondary electrons, since the electrons cause negative current relative to the beam off phase.

This test also suggests that the measurement should be done with small (few hundred volts) negative biasing. 

\subsection{Fast switching test \label{subs:fastswitch}}
\begin{figure}
\begin{centering}
\includegraphics[width=\linewidth]{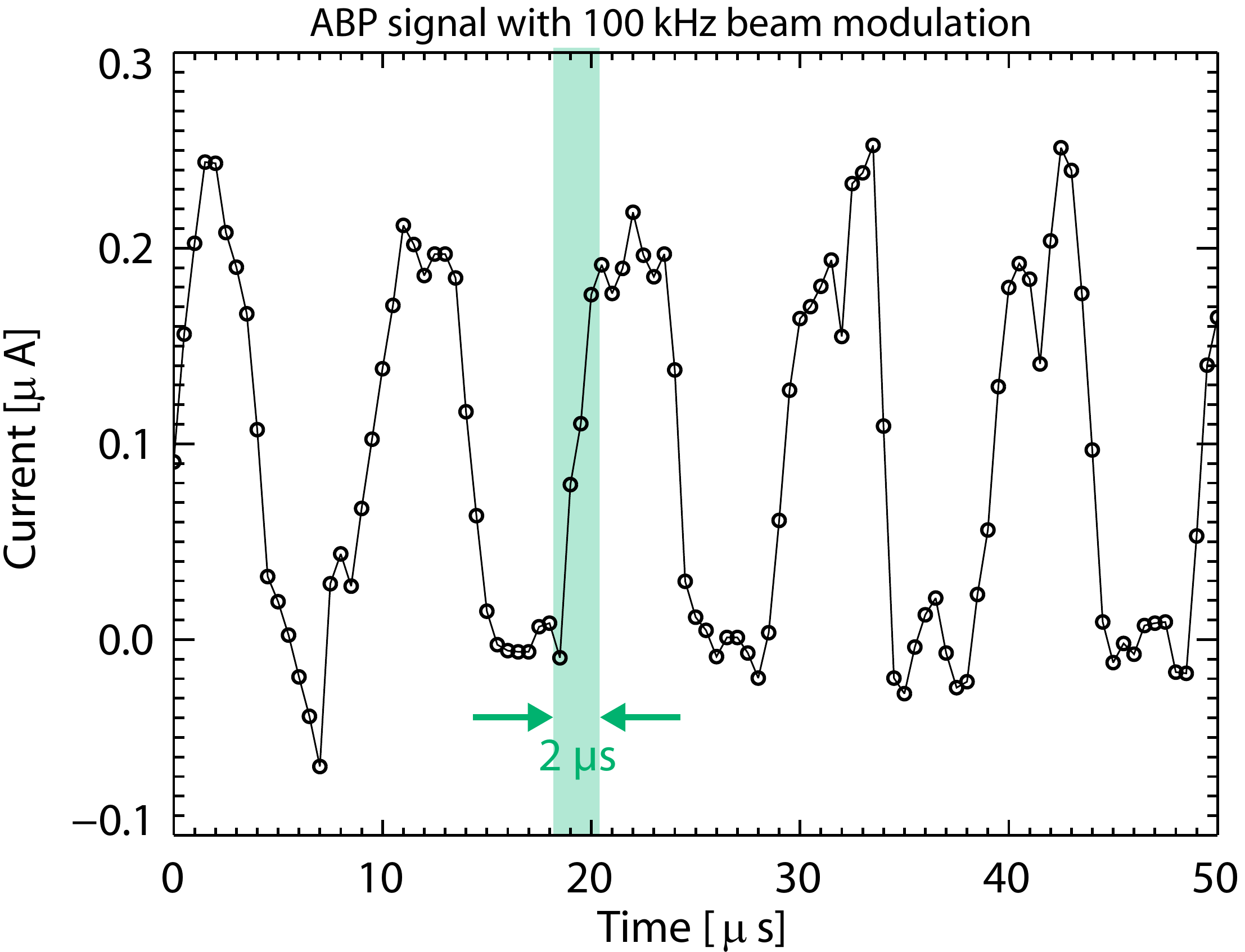} 
\par\end{centering}
\caption{Raw signal of an ABP channel during 100 kHz beam modulation.
\label{fig:fastchop}}
\end{figure}

The aim of this measurement was to clarify the change-over time between the beam on and beam off periods when fast, 100 kHz chopping is applied. This fast beam chopping will be essential in the tokamak experiment to differentiate between ion beam and background signal. The measurement was running with 2 MHz sampling, the mask was biased with -300 V. The raw signal of a channel is plotted in Figure \ref{fig:fastchop} without integration for a 50 $\mu s$ long data set (5 chopping periods) showing each Analog-to-Digital Converter (ADC) sample. The analogue bandwidth is 1 MHz, however the beam change over time is approximately 2  $\mu s$ (4 samples). This is a result of the integration time introduced by the cable capacitance and the overload protection resistor in the input of the amplifier. However, this test confirms that background can be measured on the  10 $\mu s$ timescale.

\subsection{Position sensitivity test \label{subs:fluct}}

\begin{figure}
\begin{centering}
\includegraphics[width=\linewidth]{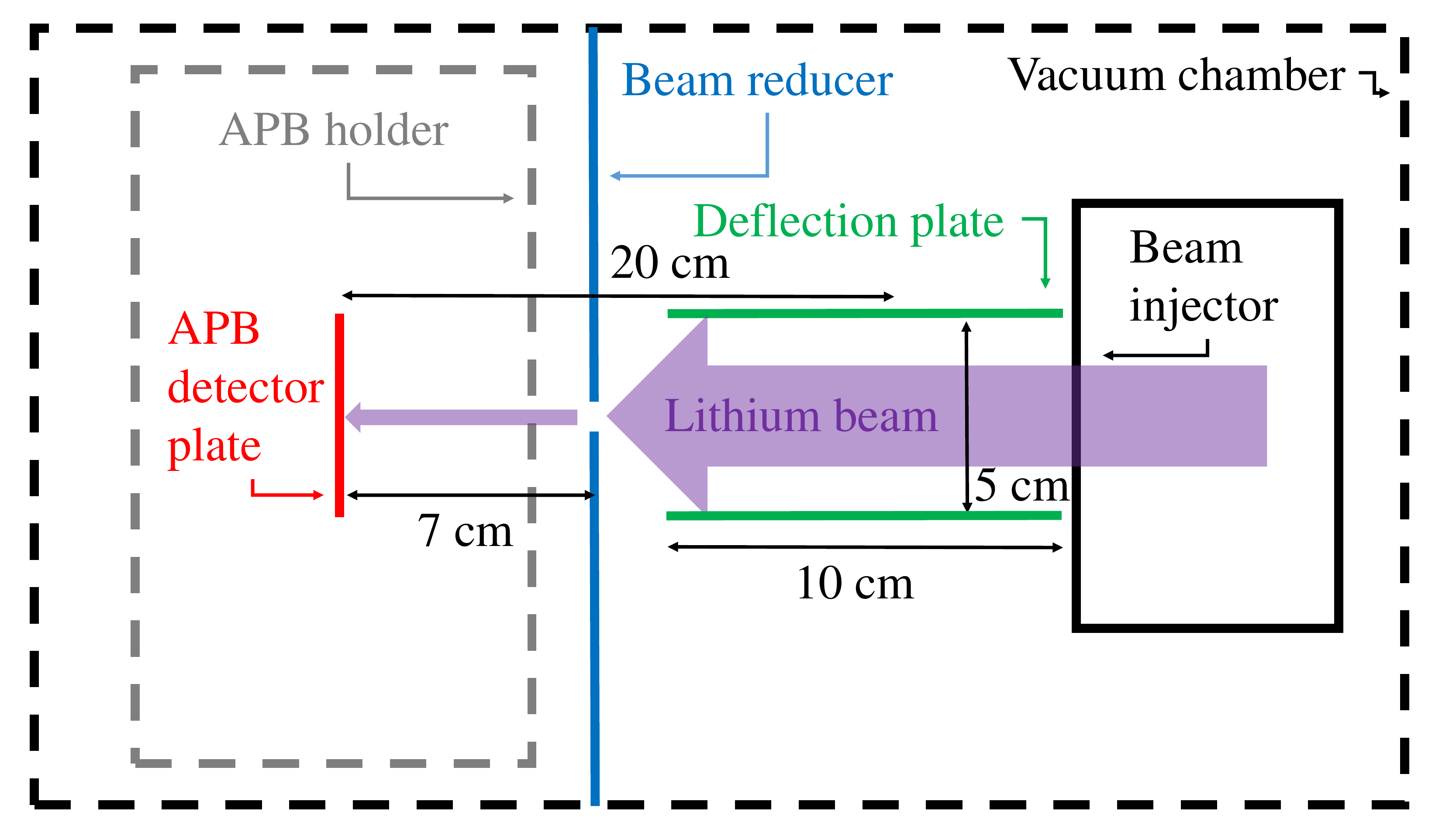} 
\par\end{centering}
\caption{Laboratory diagnostic setup with ABP placed close to the ion source \label{fig:diagsetupblock4}}
\end{figure}

The aim of this test is to characterize the sensitivity of the measurement for ion beam movement. The idea was to place the detector close to the deflection plates, so that the ion beam deflection can be reduced to 0.1 mm. 
The measurement setup was changed in the way, that the ABP detector was placed close to the ion gun, and a 5 mm diameter beam reducer diaphragm was also installed as indicated with a blue line in Figure \ref{fig:diagsetupblock4}. The angle of the beam deflection was calculated from the deflection voltage assuming homogeneous deflecting electric field between the plates. With different deflection voltages applied, the beam travels from the hole to the detector at different angles, thus the beam moves on the detector surface. Due to deflection the beam intensity also changes slightly as different areas of the 2 cm wide beam pass through the diaphragm. Please note, that the deflection plates are installed in about 30 degrees angle relative to horizontal due to technical reasons, thus the beam movement is oblique accordingly.

\subsubsection{Slow modulation test\label{subs:slowswitch}}

A 20 keV beam with approximately 0.3 mA extracted ion current was used for this measurement. The beam was deflected with 100 Hz square signal, the amplifier signal was digitized with 100 kHz sampling for 3 s, and the middle mask was biased with -300 V. 
The deflection voltage was set to 0V, 500V and 900V. 
The beam was turned off after 2.2 s to have a background measurement as well. The ion beam distribution was measured on 20 channels in a 4 $\times$ 6 channel part of the detector matrix, missing the 4 corner channels, as only a 20 channel analogue amplifier was available. Out of these 2 channels were broken at the detector plate connector (1-5, 3-5 counting from the top left corner) offering electronic background measurement for the whole measurement chain.

\begin{figure}
\begin{centering}
\includegraphics[width=\linewidth]{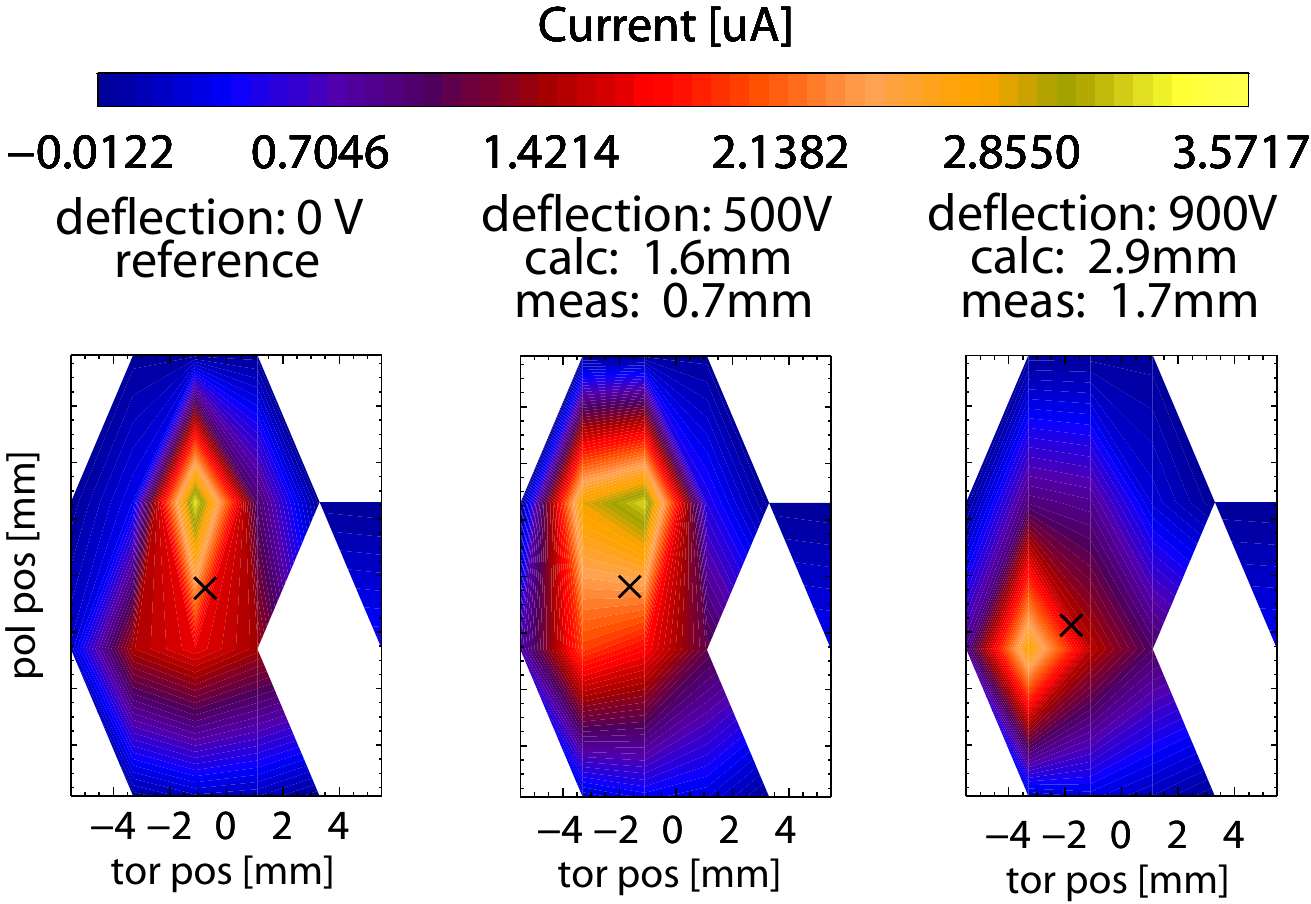} 
\par\end{centering}
\caption{Beam intensity distribution on the detector head for different deflection voltages. The bottom-right figure shows the reference distribution without deflection. Center of mass of the distributions are indicated with $\times$ sign.\label{fig:deflslowcontour}}
\end{figure}

The subfigures in figure \ref{fig:deflslowcontour} shows the ion beam current distribution on the detector for different deflection voltages, the dead channels are left blank. The bottom right subplot shows the case without deflection while the others with different deflection voltages. The calculated beam deflection and the shift in the distribution center of mass is printed above each plot. The center of mass of each distribution is indicated with $\times$ sign. The beam is deflected towards the bottom left corner relative to the undeflected case, and the effect of the increasing deflection voltage is clearly visible. The measured beam deflection is about half of the calculated if the calculated deflection is above 1 mm. This reduction in sensitivity is caused partly by the simplified beam  deflection calculation assuming homogeneous electric field between the plates, partly by the limited number of measurement channels. Nevertheless, it is confirmed that the Farady cup matrix is sensitive to sub-pixel beam movements. 
 
\begin{figure}
\begin{centering}
\includegraphics[width=\linewidth]{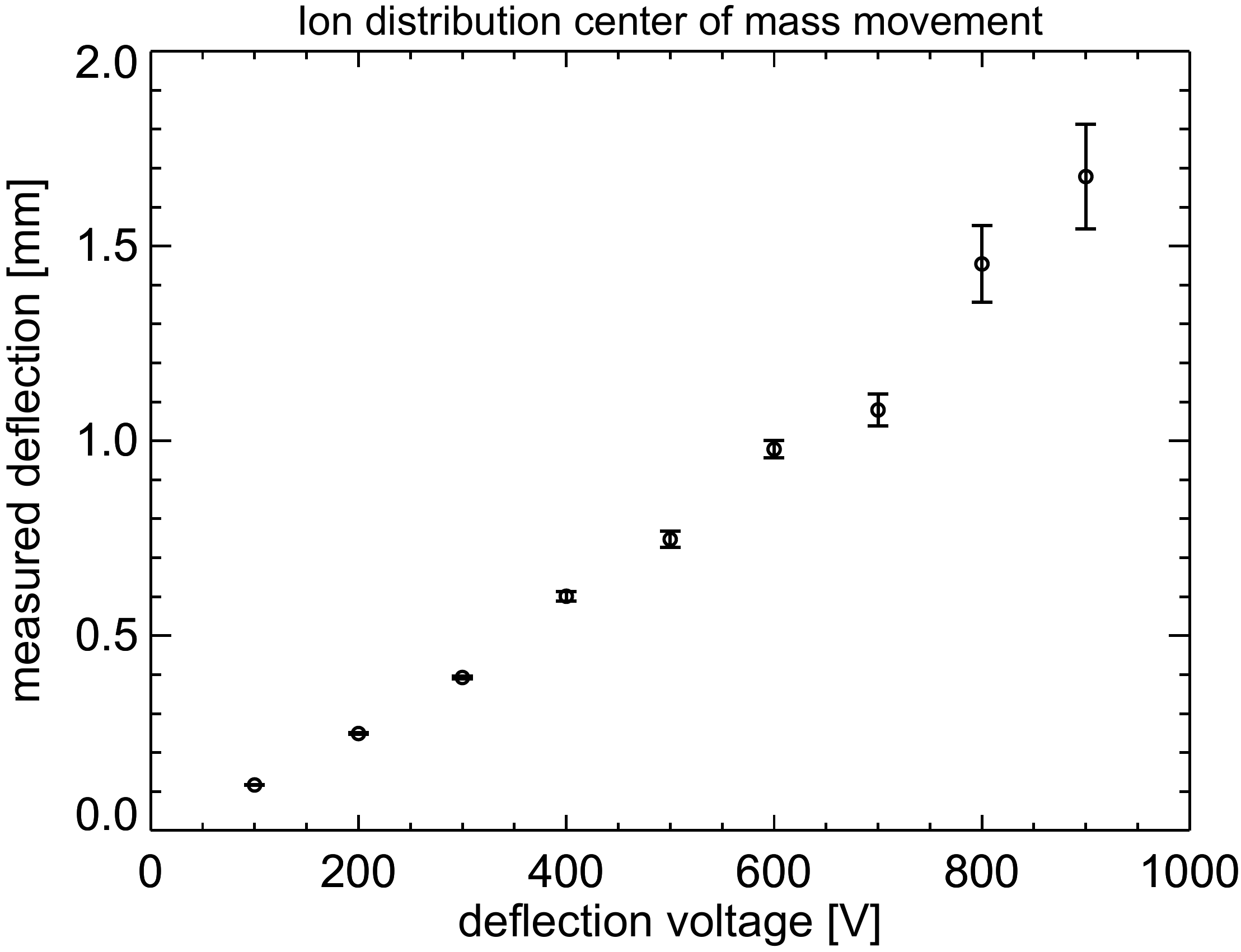} 
\par\end{centering}
\caption{Average displacement of the center of mass of the ion current distribution during the slow, 100 Hz deflection measurement for different deflection voltages. \label{fig:deflcogvsvoltageslow}}
\end{figure}

The average movement of the center of mass relative to the reference 0 V case is shown in Figure \ref{fig:deflcogvsvoltageslow} as the function of the deflection voltage. The effect of the change in the chopper voltage on the center of mass movement is close to be linear, and an order of 0.1 mm displacement is resolvable with this technique.


\subsubsection{Fast modulation test}

The aim of this test was to verify that fast (microsecond scale) beam movements can be measured.
The beam properties, the mask biasing and the deflection voltage steps were identical with the previous measurement's, while the beam modulation was set to 100 kHz and the sampling to 1 MHz. 

\begin{figure}
\begin{centering}
\includegraphics[width=\linewidth]{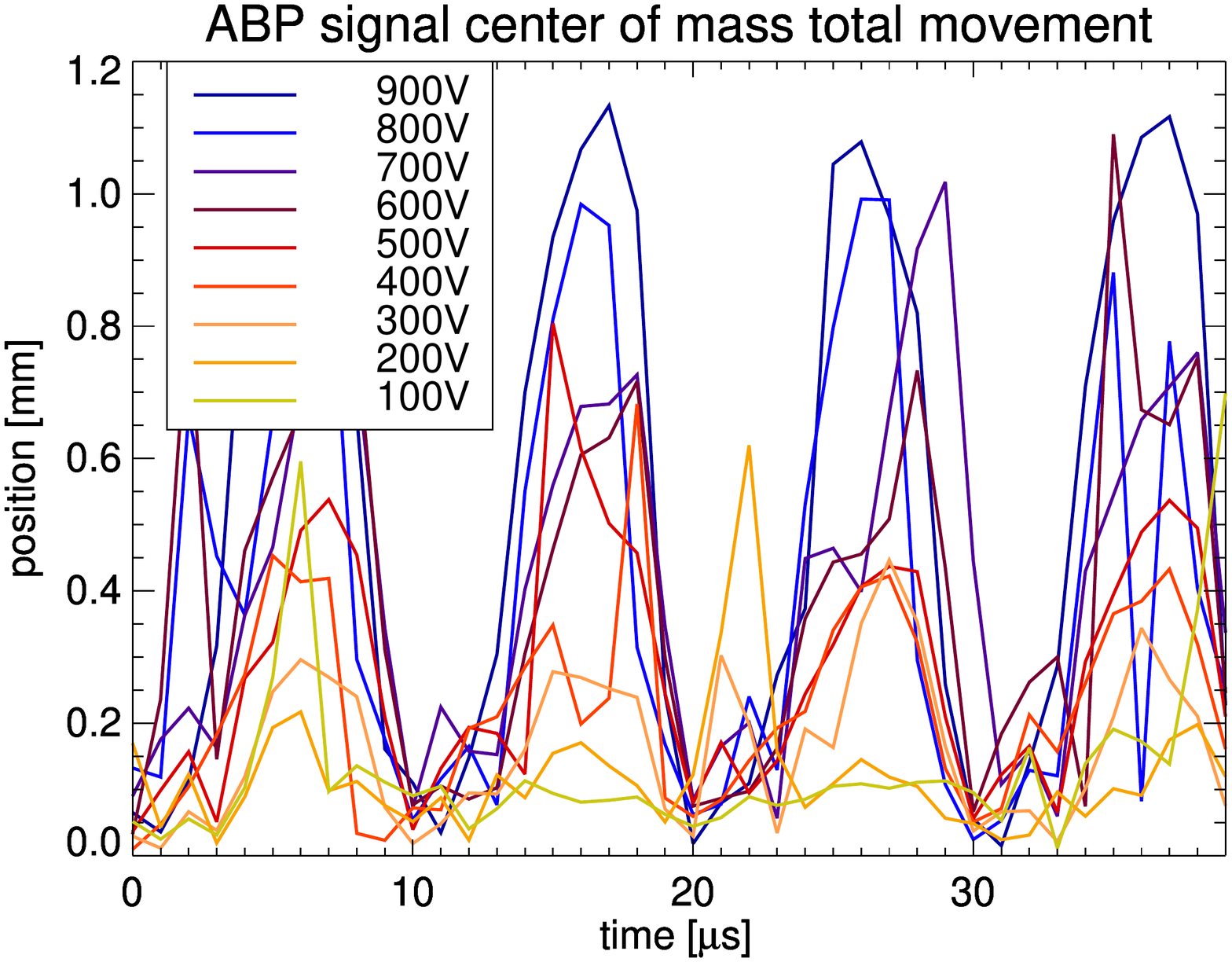} 
\par\end{centering}
\caption{Movement of the center of mass of the ion current distribution during the fast, 100 kHz deflection measurement for different deflection voltages. The different colours correspond to different deflection voltages indicated in the legend. \label{fig:deflcogfast}}
\end{figure}

The center of mass movement relative to the start point is shown in Figure \ref{fig:deflcogfast} with different colours for different deflection voltages without integration. The modulation of the center of mass is above the noise level and fairly linear. Approximately 0.2 mm movement on the microsecond time scale is resolvable.

\section{Tokamak experiment setup}\label{Sect.tokexp}

As described above, the ABP detector head has to be placed in the vacuum chamber of the torus close to the LCFS. This section summarizes the considerations for the mechanical and electrical eningeering topics and the data acquisition system for the tokamak environment.

\subsection{Detector head design}
\begin{figure}
\begin{centering}
\includegraphics[width=\linewidth]{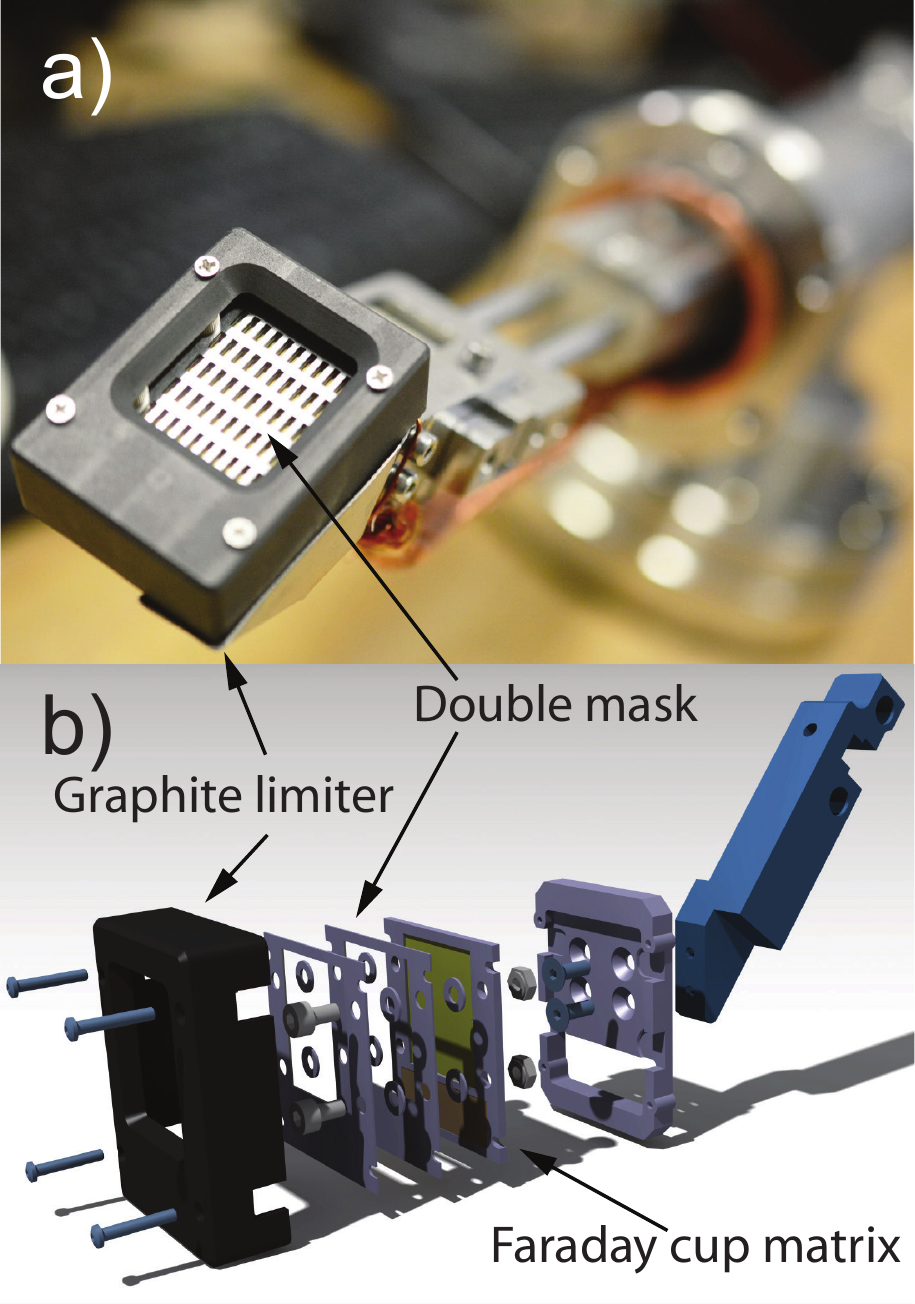} 
\par\end{centering}
\caption{Photo (a) and CAD drawing (b) of the detector head for the tokamak measurements. The detector surface is barely visible behind the graphite limiter and the double mask. \label{fig:detectorhead2}}
\end{figure}

Figure \ref{fig:detectorhead2} shows the final detector head which is installed in the COMPASS tokamak since February 2018. The most conspicuous parts are the Graphite frame and the double mask in the picture along with the actuator mechanism above, while the CAD drawing shows the inner structure of the final detector head design. The grounded graphite frame acts as a local limiter and prevents direct contact of the plasma to the masks. The actuator mechanism enables moving the detector a few cm out from the tokamak port and also 2 cm to the side.

The Faraday cups and the detector masks are connected to a vacuum feedthrough with a Kapton shielded ribbon cable. The cables are hard soldered to the connector panel on the detector board as shown in Figure~\ref{fig:detectorhead}.

\subsection{Detector holder}

The aim of the detector holder setup is to be able to move the detector, since the detailed modelling of different plasma scenarios, beam energy and beam species showed, that the detector position must be variable both vertically and horizontally to match the ion trajectories. It also aims to carry the ABP signals from the detectors to the data acquisition system, and to move the detector into the port to minimize deterioration when the detector is out of use.

To fulfill these requirements, an in-vessel setup was designed and it consists of the detector head, a horizontal actuator, a vertical actuator, an extention vacuum vessel, special cables, a double connector feedthrough and a manual actuator with proper scale to ensure position reproducibility.

\begin{figure}
\begin{centering}
\includegraphics[width=\linewidth]{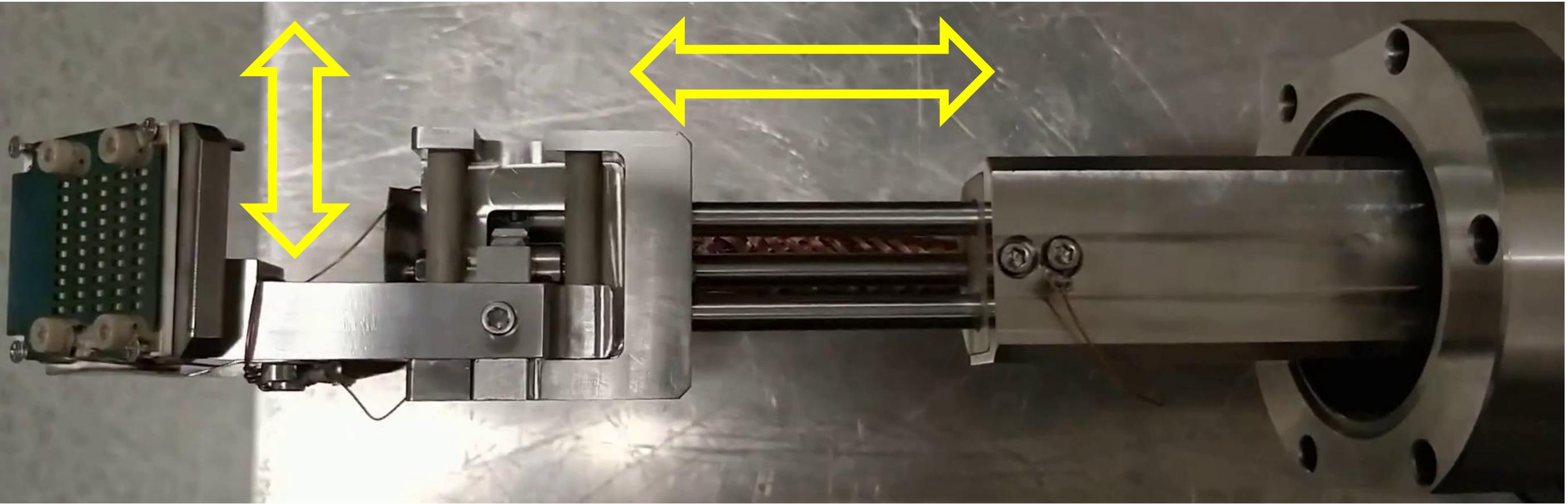} 
\par\end{centering}
\caption{Detector head on the left, and the actuator mechanics. The yellow arrows indicate the horizontal and the vertical movement capabilities respectively.  \label{fig:detectormovement}}
\end{figure}

Figure \ref{fig:detectormovement} shows the actuation possibilities with the setup, note that the picture is rotated by 90 degrees relative to its position in the tokamak. The movement range vertically is 218 mm to 273 mm in height above midplain coordinates, while horizontally 0 to 22 mm counter clockwise looking from the top of the tokamak, relative to the beam injection position.

\begin{figure}
\begin{centering}
\includegraphics[width=\linewidth]{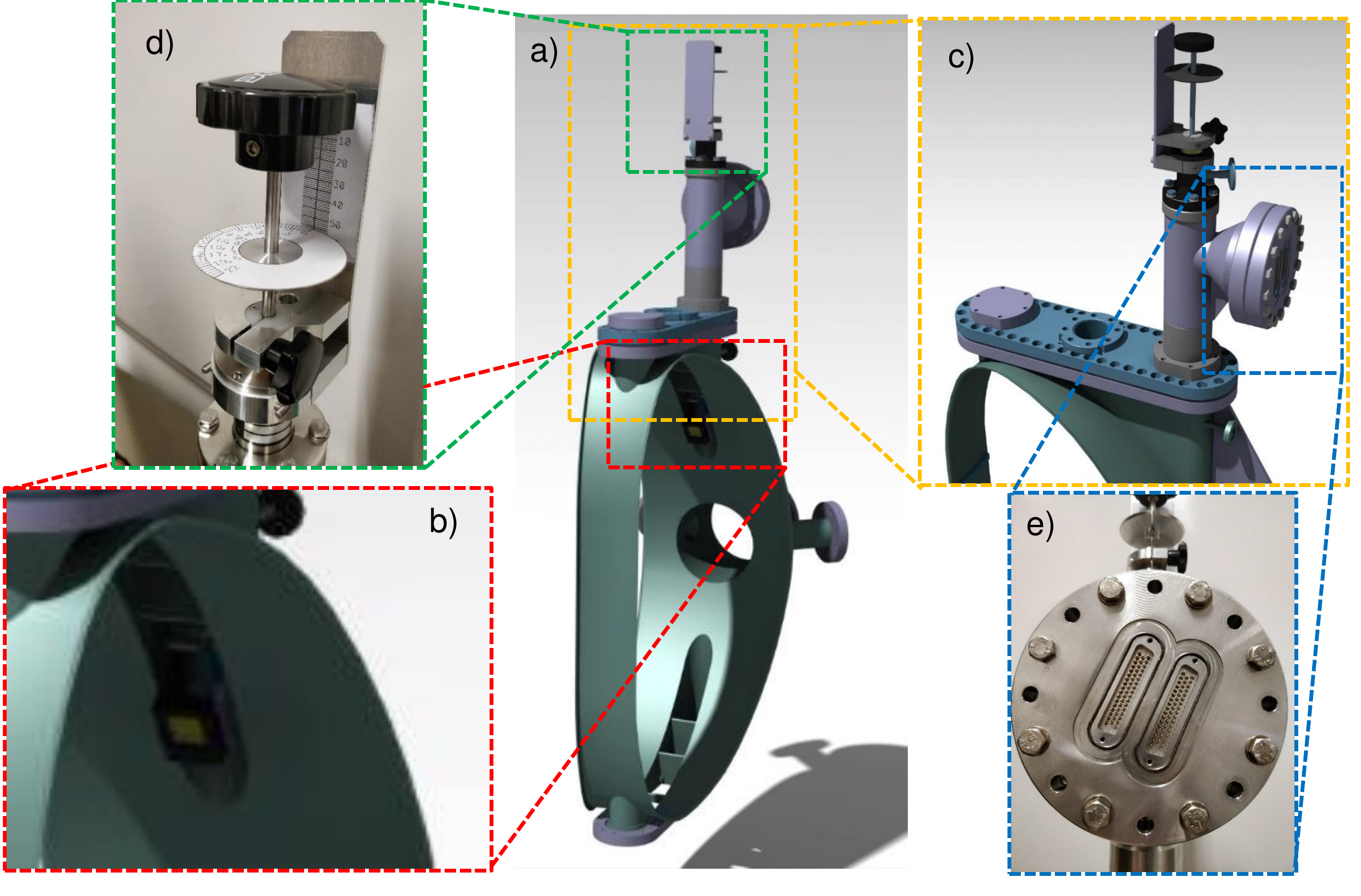} 
\par\end{centering}
\caption{A CAD drawing of a cut of the COMPASS tokamak vessel with the ABP detector setup on top can be seen in picture (a). Picture (b) shows the detector at its lowermost position in the port. Picture (c) shows the detector vacuum vessel from a different angle, while picture (d) shows the actuator mechanics, and picture (e) the vacuum feedthrough.   \label{fig:detectorholdertokamak}}
\end{figure}

Figure \ref{fig:detectorholdertokamak} indicates how the detector holder is installed at the COMPASS tokamak, picture (a) shows the CAD model of the COMPASS vacuum vessel with the ABP setup on the top. Picture (b) shows the detector head at its lowermost position, while the extention vacuum vessel and the actuator can be seen in picture (c) along with a photo of the actuator mechanics (d) and the vacuum feedthrough (e).

\section{Summary and conclusions}\label{sect:summary}

A purpose designed experimental setup and measurement series were carried out to characterize the performance of the Faraday cup type atomic beam probe diagnostic. It was found, that the Faraday cup matrix  detector with double mask is capable of measuring the expected $\sim$100 nA ion current with microsecond time resolution. A double mask is needed in front of the cups to shield the gaps between the cups so as to reduce secondary electron generation. The first mask has to be on ground potential while the second one biased to about -300 V voltage to prevent cross talk and suppress secondary electrons. Sensitivity to  sub-mm beam movement with few microsecond time resolution was confirmed, therefore this detector is a viable solution for the ABP diagnostic. Such a device has been installed on the COMPASS tokamak and its first results will be reported in a separate paper. 

\begin{acknowledgments}

This work received funding from MEYS Project No.LM2015045.

This work has been carried out within the framework of the EUROfusion Consortium and has received funding from the Euratom research and training programme 2014-2018 and 2019-2020 under grant agreement No 633053. The views and opinions expressed herein do not necessarily reflect those of the European Commission.

\end{acknowledgments}

\bibliographystyle{apsrev4-1}
\bibliography{article}

\end{document}